\documentclass[slac_one]{revtex4}
\usepackage{graphicx}
\usepackage{fancyhdr}
\begin{document}

\title{Next-to-the-Leading-Order Hadronic Corrections for Standard Model
and Beyond}

\author{Aleksandrs Aleksejevs}

\affiliation{Division of Science, SWGC, Memorial University,
Corner Brook, NL, Canada}

\author{Svetlana Barkanova}

\affiliation{Physics Department, Acadia University, Wolfville, NS, Canada}

\begin{abstract}
We explore various extensions of computational packages
such as \emph{Feynarts} and \emph{FormCalc} in application to calculation of the electron-proton scattering asymmetries relevant to the G0 and $Q_{Weak}$ experiments. Our calculations where completed up to the Next-to-the-Leading-Order (one-loop) using the dipole form factor in hadronic vertices. We saw no significant difference between monopole and dipole type of asymmetry over the range of $Q^{2} = 0.1 - 1.0\, GeV^{2}$. Our results are in the good agreement with theoretical predictions by the G0 collaboration, but show discrepancy from the experiment. This discrepancy from the experimental results is due to the strange content of the proton and possibly new physics as well.
\end{abstract}

\maketitle

\thispagestyle{fancy}

\section{Introduction}

Next-to-the-Leading-Order (NLO) effects in electroweak interactions
play a crucial role in the tests of the Standard Model and require
careful theoretical evaluation. With recent developments in the automatization
of the NLO calculations, it is reasonable to consider the possibility
that these methods could also be applied to the hadronic electro-weak
sector and thus allow for a broader application of the theory. Use
of the packages LoopTools \cite{FF,LoopTools} and Form \cite{Form} with extended packages FeynArts \cite{FeynArts}
and FormCalc \cite{FormCalc} created a possibility to perform calculations of parity-violating
asymmetries up to NLO level in electron-proton (e-p) scattering \cite{Aleks} and
hence provided us with some information on the electro-weak structure
of the nucleon. In the work presented here, we adopt FeynArts and
FormCalc for the NLO symbolic calculations of amplitude or differential
cross section in electron-nucleon scattering. Using Dirac and Pauli
type couplings with the monopole and dipole form factor approximation,
we construct the computational model enabling FeynArts and FormCalc
to deal with electron-nucleon scattering up to NLO level.  In the presented article we start with a description of the model and
then show some results relevant to G0 experiment for the momentum
transfers up to $1.0\, GeV^{2}$. In addition the prediction for e-p asymmetry
for the $Q_{Weak}$ experiment is given, and analysis of impact of
Hard-Photon-Bremsstrahlung (HPB) at low momentum transfers is provided.

\section{Main Ideas}

For the parity violating electron-proton scattering, the interaction between
Gauge bosons and proton is described in the well known form using
Dirac and Pauli form factors, and then programmed as an extension for
the FeynArts and FormCalc. To accommodate the nucleon structure, we have
the couplings preserve their vector and vector-axial structure, but
with the charges replaced by the corresponding form factors. The electromagnetic
$\Gamma_{\mu}^{\gamma-N}$ coupling has two vector components: 
\begin{equation}
\Gamma_{\mu}^{\gamma-N}\left(q\right)=ie\left[F_{1}\left(q\right)\gamma_{\mu}+\frac{i}{2m_{N}}\sigma_{\mu\alpha}q^{\alpha}F_{2}\left(q\right)\right],\label{e1.5}
\end{equation}
 where $F_{1}\left(q\right)$ and $F_{2}\left(q\right)$ are the Dirac
and Pauli form factors, respectively, and $q^{\alpha}$ is the four-momentum
transferred to the nucleon. As for $\Gamma_{\mu}^{Z-N}\left(q\right)$,
we have: 
\begin{equation}
\Gamma_{\mu}^{Z-N}\left(q\right)=ie\left[f_{1}\left(q\right)\gamma_{\mu}+\frac{i}{2m_{N}}\sigma_{\mu\alpha}q^{\alpha}f_{2}\left(q\right)+g_{A}\left(q\right)\gamma_{\mu}\gamma_{5}\right],\label{e1.6}
\end{equation}
 with $f_{1}\left(q\right)$, $f_{2}\left(q\right)$ and $g_{A}\left(q\right)$
as weak Dirac, Pauli, and axial-vector form factors. The form factors
$f_{1}\left(q\right)$ and $f_{2}\left(q\right)$ are expressed in
the following form: 
\begin{equation}
f_{1,2}\left(q\right)=\frac{1}{4cos_{W\,}sin_{W}}\left(F_{1,2}^{V\left(N\right)}\left(q\right)-4sin_{W}^{2}F_{1,2}\left(q\right)\right),\label{e1.7}
\end{equation}
 where $F_{1,2}^{V\left(p\right)}=-F_{1,2}^{V\left(n\right)}=F_{1,2}^{p}-F_{1,2}^{n}$, and
$g_{A}^{p}\left(q\right)=-g_{A}^{n}\left(q\right)=g_{A}\left(q\right)$
is an axial form factor. Our analytical expressions can be simplified considerably if we start with the monopole form factor expressed as 
\begin{equation}
\left\{ F_{1,2}\left(q\right),g_{A}\left(q\right)\right\} =\frac{\Lambda^{2}}{\Lambda^{2}-q^{2}}\left\{ F_{1,2}\left(0\right),g_{A}\left(0\right)\right\} .\label{e1.9}\end{equation}
The value of the parameter $\Lambda^{2}=0.83m_{N}^{2}$ is found from
the fit of the electromagnetic form factors by monopole approximation
in the low momentum transfer region. In order to produce e-p scattering asymmetries
for the dipole form factor, we differentiated the monopole-type asymmetries
numerically using the following prescription:\textbf{\begin{eqnarray}
\left(\frac{\Lambda^{2}}{\Lambda^{2}-q^{2}}\right)^{2} & = & -\Lambda^{4}\frac{\partial}{\partial\left(\Lambda^{2}\right)}\left(\frac{1}{\Lambda^{2}-q^{2}}\right)\label{e2.0}\end{eqnarray}
}
This way we were able to extend FeynArts and FormCalc by including
Dirac and Pauli form factors taken in the form of dipole/monopole approximation. 
Calculations were done in the on-shell renormalization scheme using the Feynman Gauge.To avoid the infrared divergences, we have treated the final asymmetries
with both Soft and Hard Photon Bremsstrahlung (SPB + HPB) contributions. Finally, it is important to point out that part of the coupling responsible for the strange content of the proton was excluded from our calculations. Hence, the discrepancy between theoretical and experimental results could be explained by the strange content of the proton.

\section{Results}

\begin{figure*}[h]
\centering
\includegraphics[width=150mm]{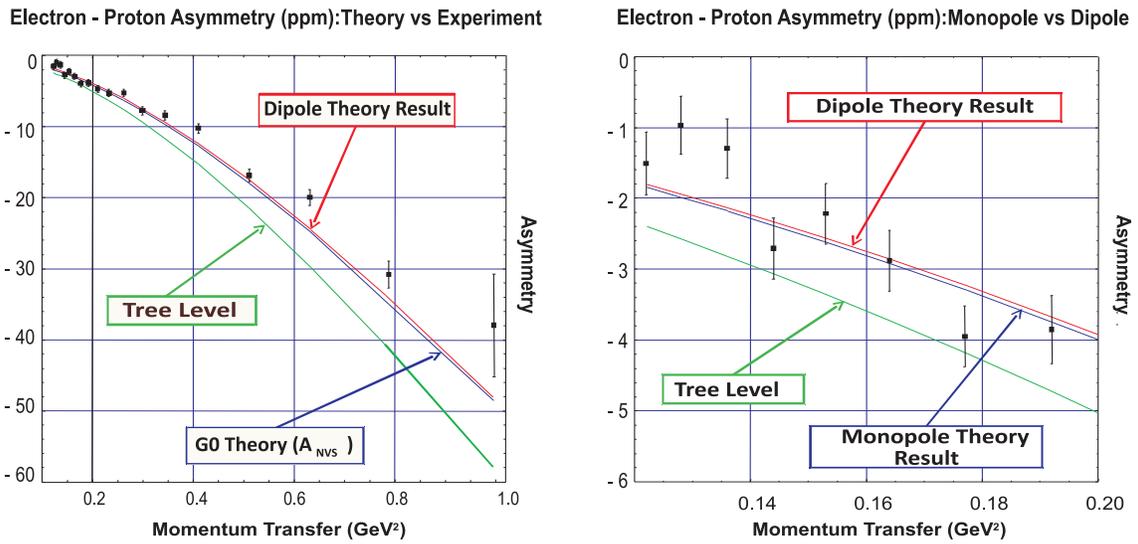}
\caption{Electron-Proton Asymmetry: a) Theory versus Experiment, b) Monopole versus Dipole.}
\label{fig:1}
\end{figure*}

A starting point for our calculations was to produce semi-automatic
calculations of asymmetries. Over the range of the momentum transfers
$Q^{2}=0.1 - 1.0\,GeV^{2}$, it was possible to compare our results to
the G0 experiment as well as to the theoretical results of no-vector-strange
asymmetry ($A_{NVS}$ ) taken from G0 collaboration \cite{G0}.\textbf{
}As it can be seen from Fig.(\ref{fig:1}a), our results are very close to the results of G0, but there is some discrepancy between the theory and experiment, which is most probably due to the strange content of the proton.%

At the momentum transfers over the range of $Q^{2} = 0.1-1.0\,GeV^{2}$, 
we saw no distinguishable difference between monopole and dipole type
of results. To illustrate impact of this difference, we changed the
scale to $Q^{2}=0.1-0.2\,GeV^{2}$ (see Fig.(\ref{fig:1}b)) so it is evident that our calculations are independent from the
choice of the model for the form factors. The negligible form factor model
dependence can be explained by the relevant contributions coming from the boxes
and proton vertex corrections graphs only.

\begin{figure*}[h]
\begin{centering}
\includegraphics[width=75mm]{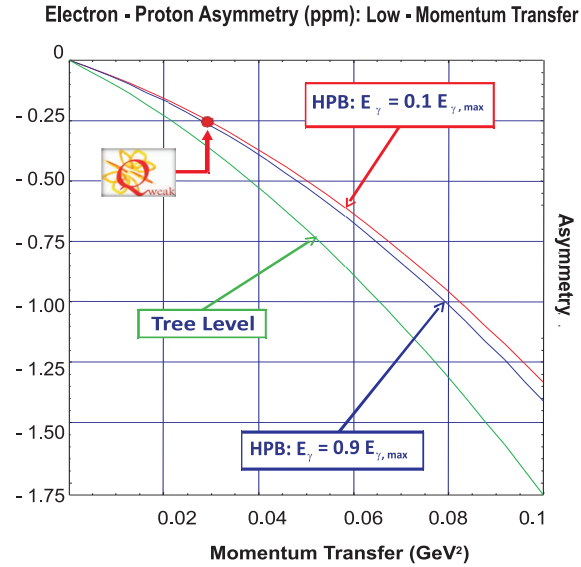}
\caption{Electron-Proton Asymmetry: Low-Momentum transfer.}
\label{fig:2}
\par\end{centering}
\end{figure*}

For kinematics relevant to the $Q_{Weak}$ experiment \cite{Qweak}, our prediction for
the asymmetry was\textbf{\begin{equation}
A_{Q_{Weak}}(Q^{2}=0.03\, GeV^{2})=-0.265\pm0.007\, ppm,\label{e2.1}\end{equation}
}with small uncertainty primarily coming from the different cuts
on energy of the emitted HPB photons. Although HPB cuts have big impact
on the asymmetry at high momentum transfer, it is quite negligible for
the low momentum transfers (See Fig.(\ref{fig:2})). 

Comparing the weak charge of the proton extracted from our asymmetry predictions
with upcoming $Q_{Weak}$ experimental result may potentially open a
window for the new physics. Our plans include development and use of the advanced symbolic computational
methods to study various production and decay channels in hadronic
physics with the new degree of precision.

\end{document}